# A Locally Deterministic, Detector-Based Model of Quantum Measurement

**Brian R. La Cour**

**Abstract** This paper describes a simple, causally deterministic model of quantum measurement based on an amplitude threshold detection scheme. Surprisingly, it is found to reproduce many phenomena normally thought to be uniquely quantum in nature. To model an $N$-dimensional pure state, the model uses $N$ complex random variables given by a scaled version of the wave vector with additive complex noise. Measurements are defined by threshold crossings of the individual components, conditioned on single-component threshold crossings. The resulting detection probabilities match or approximate those predicted by quantum mechanics according to the Born rule. Nevertheless, quantum phenomena such as entanglement, contextuality, and violations of Bell's inequality under local measurements are all shown to be exhibited by the model, thereby demonstrating that such phenomena are not without classical analogs.

## 1 Introduction

Quantum mechanics exhibits many peculiar and surprising phenomena. Indeterminism, wave/particle duality, entanglement, contextuality, and nonlocality are just a few of the more salient examples. Past attempts to provide a realistic local and deterministic interpretation of quantum phenomena have been frustrated by the many "no go" theorems ruling out various classes of hidden variable theories [1,2]. Most notable among these are the Kochen–Specker theorem [3], which rules out non-contextual

B. R. La Cour (✉)
Applied Research Laboratories, The University of Texas at Austin, P.O. Box 8029, Austin,
TX 78713-8029, USA
e-mail: blacour@arlut.utexas.edu

hidden variable models, and Bell's theorem [4], which addresses local hidden variable models. This paper offers a simple mathematical model that exhibits many of these distinctly quantum phenomena, thereby demonstrating that they are not uniquely quantum in nature. Indeed, many apparently quantum phenomena have found analogs in classical mechanics and even social science [5,6]. Although rather simple in its present form, it is hoped that this model may form the basis of a more sophisticated physical theory. Such a model may also help to elucidate whether an epistemic or ontic interpretation of the quantum state is more appropriate [7,8].

The proposed model can be described succinctly as a complex random vector $\boldsymbol{a}$ for which measurements consist of threshold crossings of component magnitudes. The specific form is that of a fixed "classical" signal plus a random "quantum" noise term, defined as follows. Suppose we wish to model a given "design" state vector $|\psi\rangle$ in an $N$-dimensional Hilbert space. The components in some standard basis are given by $\alpha_n := \langle n | \psi \rangle$ for $n = 1, \ldots, N$. The aforementioned random vector $\boldsymbol{a}$ is then defined to be

$$\boldsymbol{a} := s\boldsymbol{\alpha} + \boldsymbol{w}, \quad (1)$$

where $\boldsymbol{\alpha} = [\alpha_1, \ldots, \alpha_N]^\mathsf{T} \in \mathbb{C}^N$ is the normalized signal and $s \geq 0$ is the signal amplitude. Normalization is in the usual sense that

$$\|\boldsymbol{\alpha}\|^2 := \sum_{n=1}^N |\alpha_n|^2 = 1. \quad (2)$$

The term $\boldsymbol{w} = [w_1, \ldots, w_N]^\mathsf{T}$ is defined as a complex random noise vector whose joint distribution will be specified later. Either $\boldsymbol{w}$ or $\boldsymbol{a}$ may be construed as the "hidden variable" whose specific realization determines the outcome of a measurement.

The physical motivation for this model stems from early work in stochastic electrodynamics (SED) [9]. In SED, quantum phenomena are hypothesized to arise from classical interactions of matter with a real, albeit stochastic, background electromagnetic field corresponding to the (virtual) vacuum field of quantum electrodynamics. Stochastic optics (SO), a natural extension of SED, attempts to use this same hypothesis to explain phenomena in quantum optics [10]. In the view of SO, the underlying reality corresponding to a quantum state $|\psi\rangle$ is a real (as opposed to virtual) electromagnetic field (e.g., a plane wave of a particular mode), what one might call the "signal" that the experimenter prepares, plus a stochastic background component corresponding to what one might call the "noise" of the vacuum field. From this viewpoint, one may interpret $\boldsymbol{a}$ as giving the amplitudes and phases of the $N$ modes of a classical electromagnetic field.

While providing an intuitively appealing and qualitatively accurate description of many quantum optical phenomena, SO has also made predictions at variance with experiment [11,12]. More relevant this paper, however, is its lack of deterministic outcomes. In SO, detector probabilities are determined semiclassically as a function of mode intensity. Given a particular realization, only the probability of a detection is specified, not the outcome. In the model proposed here, we shall recover determinism by defining detections as amplitude threshold crossings. Thus, given a threshold $\gamma \geq 0$, we shall say that a measurement in the standard basis results in outcome $n$ if and only if $|a_n| > \gamma$ and $|a_{n'}| \leq \gamma$ for all $n' \neq n$. Instances of $\boldsymbol{a}$ for which there are no threshold crossings are rejected as "non-detections." Likewise, instances of more then one threshold crossing are rejected through post-selection.

The restriction to single-detection events may seem artificial, but it corresponds to what is commonly done in the laboratory. In quantum optics, for example, single photon production rates can be quite small [13]. Multiple detections are even rarer, and dark counts may occur even when there is no signal (corresponding to $s = 0$ in the present model). In such experiments, one often works in the so-called "coincidence basis," in which a detected and heralding photon are coincidently observed. This corresponds operationally to the aforementioned post-selection procedure. As will be shown later, this conditioning is key to reproducing many important quantum phenomena [14].

A final point to be defined in the model regards unitary transformations and measurements in other bases. In the proposed model, a unitary operator $U$ which transforms $|\psi\rangle$ to $U|\psi\rangle$ similarly transforms $\boldsymbol{a}$ to $U\boldsymbol{a}$ for a given realization of $\boldsymbol{a}$. Thus, if we measure an observable for which $U$ is a diagonalizing unitary matrix of eigenvectors, then $U^\dagger \boldsymbol{a}$ is used in place of $\boldsymbol{a}$ to determine the measurement outcome. Since one may, in this manner, unambiguously determine the outcome that would have been obtained had a different observable been chosen, it follows that the model is not only deterministic but counterfactually definite [15]. As we shall see, this property plays a key role in understanding contextuality and quantum nonlocality.

Interestingly, a similar threshold-based quantum measurement scheme has recently been developed by Khrennikov [16], who uses a model of a complex, vector-valued stochastic process $\{\boldsymbol{\phi}(t) : t \in \mathbb{R}\}$. This process is assumed to be zero-mean and have a covariance of $E[\phi_i(s)^*\phi_j(t)] = B_{ij}\sqrt{|st|}\,\delta(s-t)$. Detections are made when the amplitude of the process, suitably time averaged, falls above some threshold $\mathcal{E}_d$. The average time between such threshold crossings (or "clicks") for component $i$ is found to be $\bar{\tau}_i = B_{ii}/\mathcal{E}_d$. Thus, the fraction of all clicks that are from component $i$ is $B_{ii}/\text{Tr}(B)$, which is interpreted as a probability. This choice of normalization is equivalent to conditioning on single-detection events. Incorporating a background field and properly calibrating the threshold even allows violations of Bell's inequality [17]. Since classical fields are used, the model is manifestly local. It is, however, nonobjective: the values of observables cannot be assigned in advance.

The outline of the paper is as follows. A mathematical description of the detection probabilities, with some simple examples, is given in Sect. 2. This description is extended in Sect. 3 to the case of probabilities conditioned on single-detection events, wherein the Born rule is recovered under certain limiting conditions. The measurement model description is completed in Sect. 4 with a discusion of unitary transformations. A key result there is that, for certain choices of noise models, the Born rule is preserved under unitary transformations. Using this result, it is shown that certain quantum states can be deduced empirically using quantum state tomography. Quantum contextuality is studied in Sect. 5 using the example of the Mermin–Peres magic square. There, it is shown that, by conditioning on single-detection events, one is able to reproduce all quantum phenomena yet remain deterministic. The question of entanglement is taken up in Sect. 6. It is shown that the proposed model is empirically equivalent to a Bell state, as may be inferred through quantum state tomography. Furthermore, it is shown that, by conditioning on single-detection events, this model is capable of producing violations of Bell's inequality in both simultaneous and space-like separated measurements.

## 2 Detection Probabilities

Given a vector $\boldsymbol{\alpha}$ corresponding to a design state $|\psi\rangle$, let $P_n(\boldsymbol{\alpha}, \gamma)$ denote the probability (given by the distribution of $\boldsymbol{w}$, the signal amplitude $s$, and the threshold $\gamma$) that a single threshold crossing of component $n$ occurs. Similarly, let $P_0(\boldsymbol{\alpha}, \gamma)$ denote the probability that no threshold crossing occurs. Specifically,

$$P_n(\boldsymbol{\alpha}, \gamma) := \Pr\left[|a_n| > \gamma, \ |a_{n'}| \leq \gamma \ \forall n' \neq n\right] \quad (3)$$

and

$$P_0(\boldsymbol{\alpha}, \gamma) := \Pr\left[|a_1| \leq \gamma, \ \ldots, \ |a_N| \leq \gamma\right]. \quad (4)$$

The probability of obtaining more than one detection will be denoted $P_\infty(\boldsymbol{\alpha}, \gamma)$; thus,

$$P_\infty(\boldsymbol{\alpha}, \gamma) := 1 - P_0(\boldsymbol{\alpha}, \gamma) - \sum_{n=1}^{N} P_n(\boldsymbol{\alpha}, \gamma). \quad (5)$$

The condition of having at most one detection, i.e. $P_\infty(\boldsymbol{\alpha}, \gamma) = 0$, can be achieved asymptotically by choosing a sufficiently large threshold, since the probability of multiple detections will tend to zero in such a limit. It is possible, however, and may be of some utility, to achieve this condition explicitly. This may be done quite easily by normalizing the noise vector to some fixed value. Specifically, we have the following theorem.

**Theorem 1** *Suppose the noise is bounded ($\|\boldsymbol{w}\| \leq \sigma$ almost surely for some $\sigma \geq 0$), the detection threshold is sufficiently high ($\gamma \geq \sigma$), and the signal strength is sufficiently low ($s \leq (\sqrt{2} - 1)\sigma$). Then a detection occurs for at most one value of n.*

*Proof* Suppose $|a_n| > \gamma$ and $|a_{n'}| > \gamma$ for some $n \neq n'$. Then $\|\boldsymbol{a}\|^2 \geq |a_n|^2 + |a_{n'}|^2 > 2\gamma^2 \geq 2\sigma^2$, so $\|\boldsymbol{a}\| > \sqrt{2}\sigma$. But, by the triangle inequality, $\|s\boldsymbol{\alpha} + \boldsymbol{w}\| \leq s\|\boldsymbol{\alpha}\| + \|\boldsymbol{w}\| \leq (\sqrt{2} - 1)\sigma + \sigma$, so $\|\boldsymbol{a}\| \leq \sqrt{2}\sigma$. We thus arrive at a contradiction and conclude that two or more detections cannot occur. □

### 2.1 Simple Two-Dimensional Example

Suppose $N = 2$ and $|\psi\rangle = |1\rangle$, so that $\boldsymbol{\alpha} = [1, 0]^\mathsf{T}$. Let $\boldsymbol{w} = \sigma e^{i\theta}[0\ 1]^\mathsf{T}$, where $\theta$ is uniformly distributed on the interval $[0, 2\pi)$. Thus, we have the random complex amplitude vector

$$\boldsymbol{a} = \begin{pmatrix} s \\ \sigma e^{i\theta} \end{pmatrix}. \quad (6)$$

Now note that, for $\gamma = \sigma$ and $s > \sigma$, we have $P_0(\boldsymbol{\alpha}, \gamma) = 0$, $P_1(\boldsymbol{\alpha}, \gamma) = 1$, $P_2(\boldsymbol{\alpha}, \gamma) = 0$, and $P_\infty(\boldsymbol{\alpha}, \gamma) = 0$. This mimics the behavior of a single photon in a definite state (polarization and wave vector mode).

Suppose instead that the design quantum state is given by $|\psi\rangle = \frac{1}{\sqrt{2}}[|1\rangle + |2\rangle]$ and let $\boldsymbol{w} = (\sigma/\sqrt{2})e^{i\theta}[1, -1]^\mathsf{T}$, where again $\theta$ is uniformly distributed on the interval $[0, 2\pi)$. We now have

$$\boldsymbol{a} = \frac{1}{\sqrt{2}} \begin{pmatrix} s + \sigma e^{i\theta} \\ s - \sigma e^{i\theta} \end{pmatrix}. \tag{7}$$

Note that $|a_1|^2 = \frac{1}{2}|s + \sigma e^{i\theta}|^2 = \frac{1}{2}(s^2 + \sigma^2 + 2s\sigma \cos\theta)$ while $|a_2|^2 = \frac{1}{2}(s^2 + \sigma^2 - 2s\sigma \cos\theta)$. Because the noise terms are correlated via the common variable $\theta$, we see that $|a_1|$ is large when $|a_2|$ is small and vice versa. In particular, as $s$ approaches $\sigma$ from above, we see that $P_0(\boldsymbol{\alpha}, \gamma) = 0$, $P_1(\boldsymbol{\alpha}, \gamma) \to \frac{1}{2}$, $P_2(\boldsymbol{\alpha}, \gamma) \to \frac{1}{2}$, and $P_\infty(\boldsymbol{\alpha}, \gamma) \to 0$.

The probability of two simultaneous detections goes to zero in this case. Instead, a single detection occurs for either component with equal probability. This mimics the "particle-like" behavior observed in single-photon experiments. Note that Eqs. (6) and (7) are related to one another via a Hadamard transform, representing the action of a beamsplitter. This provides a physical picture relating Eqs. (6) and (7). That noise may contribute to (or, indeed, be necessary for) signal detection is a phenomenon similar to stochastic resonance [18].

Although this is a two-dimensional model, one may consider it to be four-dimensional by taking $[a_1, a_2, a_3, a_4] = [a_1, s + \sigma e^{i\theta}, s - \sigma e^{i\theta}, a_4]^\mathsf{T}/\sqrt{2}$ and $|a_1| = |a_4| \leq \sigma$. This may be construed as an entangled state of, say, a single photon with the vacuum state. Thus, even in this very simple example, we find evidence of quantum entanglement in what may be considered a classical model.

### 2.2 Another Two-Dimensional Example

Suppose again that $N = 2$ and $|\psi\rangle = |1\rangle$, so $\boldsymbol{\alpha} = [1, 0]^\mathsf{T}$. Now, however, let $\boldsymbol{w} = \sigma[e^{i\phi_1}\cos\theta, e^{i\phi_2}\sin\theta]^\mathsf{T}$, where $\phi_1, \phi_2$ are uniformly distributed on the interval $[0, 2\pi)$ and $\theta$ has the probability density function $\sin(2\theta)$ over the interval $[0, \frac{\pi}{2}]$. This corresponds to a uniform distribution over the Bloch sphere. Adding the signal, we have the random complex amplitude vector

$$\boldsymbol{a} = \begin{pmatrix} s + \sigma e^{i\phi_1} \cos\theta \\ \sigma e^{i\phi_2} \sin\theta \end{pmatrix}. \tag{8}$$

Note that, for $\gamma = \sigma$ and $s = (\sqrt{2} - 1)\sigma$, we have $P_1(\boldsymbol{\alpha}, \gamma) > 0$, $P_2(\boldsymbol{\alpha}, \gamma) = 0$ and, by Theorem 1, $P_\infty(\boldsymbol{\alpha}, \gamma) = 0$. Conditioned on there being a detection, then, the outcome will always be $n = 1$, but there is a nonzero probability of missed detections.

Suppose now the design quantum state is given by $|\psi\rangle = \frac{1}{\sqrt{2}}[|1\rangle + |2\rangle]$ and let $\boldsymbol{w}$ be as before. We now have

$$\boldsymbol{a} = \begin{pmatrix} s/\sqrt{2} + \sigma e^{i\phi_1} \cos\theta \\ s/\sqrt{2} + \sigma e^{i\phi_2} \sin\theta \end{pmatrix}. \tag{9}$$

Since $\cos\theta$ and $\sin\theta$ have the same distribution, $P_1(\boldsymbol{\alpha}, \gamma) = P_2(\boldsymbol{\alpha}, \gamma) > 0$. Conditioned on there being a detection, then, it is equally likely to be $n = 1$ or $n = 2$. Furthermore, since $P_\infty(\boldsymbol{\alpha}, \gamma) = 0$, by Theorem 1, simultaneous detections cannot occur, although the possibility remains of there being no detections at all.

Although it is not obvious, the stochastic models of Eqs. (8) and (9) are related by a Hadamard transform as well. Furthermore, it can be shown that $\boldsymbol{w}$ has the same distribution as $\sigma z/\|z\|$, where $z$ is a standard complex Gaussian random vector (i.e., with mean vector $\mathrm{E}[z] = 0$ and covariance matrix $\mathrm{E}[zz^\dagger] = I$, where $I$ is the identity and $\mathrm{E}[\cdot]$ represents an expectation). Indeed, from this very fact one can show that the two equations are so related.

This example was first introduced by Marshall and Santos [19] in the context of SO to explain certain quantum optics effects, such as the wave/particle duality of light, photon antibunching, and experimental tests of Bell's inequality. In their interpretation, $\boldsymbol{a}$ is the transverse electric field of a classical plane wave and $\boldsymbol{w}$ represents the component of that field due the zeropoint vacuum. The present model differs from that of Ref. [19] in that they assumed a detection probability of the form $P_n(\boldsymbol{\alpha}, \gamma) \propto \max(0, 2\mathrm{E}[|a_n|^2] - \gamma)$. Here, we assume only that the detection probabilities are determined by the frequency of threshold-crossing events.

## 3 Conditional Detections

Theorem 1 shows that, under suitable conditions, $P_\infty(\boldsymbol{\alpha}, \gamma) = 0$ for all $\boldsymbol{\alpha}$. A similar result may effectively be obtained by simply increasing the threshold. This has the reciprocal effect of reducing the number of single detections, of course, but we may then condition (or post-select) on just these events. Let us, then, define the conditional probability $p_n(\boldsymbol{\alpha}, \gamma)$ that a single detection of $n$ occurs as follows:

$$p_n(\boldsymbol{\alpha}, \gamma) := \frac{P_n(\boldsymbol{\alpha}, \gamma)}{P_1(\boldsymbol{\alpha}, \gamma) + \cdots + P_N(\boldsymbol{\alpha}, \gamma)} \qquad (10)$$

A key result is the following theorem.

**Theorem 2** *If $\boldsymbol{\alpha}$ is such that all nonzero components are equal in magnitude, then $p_n(\boldsymbol{\alpha}, \gamma) \to |\alpha_n|^2$ as $\gamma \to \infty$ for all $n = 1, \ldots, N$ (i.e., the Born rule holds asymptotically), provided that the statistical distribution of $\boldsymbol{w}$ has the following properties:*

1. The components of $\boldsymbol{w}$ are identically distributed.
2. $P_n(\boldsymbol{\alpha}, \gamma) > 0$ for all $\boldsymbol{\alpha}, \gamma$, and $n = 1, \ldots, N$.
3. Whenever $s > 0$ and $|\alpha_m| < |\alpha_k|$, the ratio $P_m(\boldsymbol{\alpha}, \gamma)/P_k(\boldsymbol{\alpha}, \gamma)$ tends to zero as $\gamma \to \infty$.

*Proof* Without loss of generality, suppose $\alpha_k \neq 0$ for $k = 1, \ldots, K$, where $1 \leq K \leq N$. Since the components are equal in magnitude, $\alpha_k = e^{i\theta_k}/\sqrt{K}$ for some $\theta_k \in [0, 2\pi)$. For $k, k' \leq K$, we see that $P_k(\boldsymbol{\alpha}, \gamma) = P_{k'}(\boldsymbol{\alpha}, \gamma)$, by symmetry, since $|a_k|$ has the same distribution as $|1/\sqrt{K} - w_k|$ and $w_k$ has the same distribution as $w_{k'}$. (Note that independence is not needed, only identicality.) Similarly, $P_m(\boldsymbol{\alpha}, \gamma) = P_{m'}(\boldsymbol{\alpha}, \gamma)$ for $m, m' > K$. Now consider the conditional probability $p_k(\boldsymbol{\alpha}, \gamma)$, which may be written

$$p_k(\boldsymbol{\alpha}, \gamma) = \frac{P_k(\boldsymbol{\alpha}, \gamma)}{K P_k(\boldsymbol{\alpha}, \gamma) + (N - K) P_m(\boldsymbol{\alpha}, \gamma)}$$
$$= \frac{1}{K + (N - K) P_m(\boldsymbol{\alpha}, \gamma)/P_k(\boldsymbol{\alpha}, \gamma)}. \quad (11)$$

As $\gamma \to \infty$, we see that $p_k(\boldsymbol{\alpha}, \gamma) \to 1/K$ and, similarly, $p_m(\boldsymbol{\alpha}, \gamma) \to 0$, in accordance with the Born rule. □

The class of possible distributions for $\boldsymbol{w}$ satisfying Theorem 2 is quite general. In particular, it includes the case of independent and identically distributed Gaussian noise (i.e., $\boldsymbol{w} = \sigma \boldsymbol{z}$). A proof is given in the appendix.

The theorem may be modified to apply to cases in which the noise is bounded. In particular, we have the following.

**Theorem 3** *If $\boldsymbol{\alpha}$ is such that all nonzero components have an equal magnitude of $1/\sqrt{K}$ for some $K > 0$, then, provided $\gamma < \sqrt{s^2/K + \sigma^2}$, we have $p_n(\boldsymbol{\alpha}, \gamma) = |\alpha_n|^2$ for all $n = 1, \ldots, N$ (i.e., the Born rule holds exactly), provided that the statistical distribution of $\boldsymbol{w}$ has the following properties:*

1. *The components of $\boldsymbol{w}$ are identically distributed.*
2. $P_n(\boldsymbol{\alpha}, \gamma) > 0$ *if* $\gamma < \sqrt{|s\alpha_n|^2 + \sigma^2}$.
3. $P_n(\boldsymbol{\alpha}, \gamma) = 0$ *if* $\gamma \geq \sqrt{|s\alpha_n|^2 + \sigma^2}$.

*Proof* The proof proceeds initially as for Theorem 2, with $P_k(\boldsymbol{\alpha}, \gamma) > 0$ and $P_m(\boldsymbol{\alpha}, \gamma) = 0$, provided $\sigma \leq \gamma < \sqrt{s^2/K + \sigma^2}$. It is then clear that $p_k(\boldsymbol{\alpha}, \gamma) = 1/K$, while $p_m(\boldsymbol{\alpha}, \gamma) = 0$. □

Property 3 of Theorem 3 is satisfied by any $\boldsymbol{w}$ such that $\|\boldsymbol{w}\| \leq \sigma$. Properties 1 and 2 would be satisfied, for example, by $\boldsymbol{w} = \sigma \boldsymbol{z}/\|\boldsymbol{z}\|$ with $\gamma = \sigma$ and $s = (\sqrt{2} - 1)\sigma$.

Although limited in scope with respect to the applicable values of $\boldsymbol{\alpha}$, Theorems 2 and 3 cover a broad range of interesting quantum states, including the standard basis states and several maximally entangled states, such as the Bell states, to be discussed later, and Greenberger–Horne–Zeilinger (GHZ) states [20]. Furthermore, even under conditions that do not satisfy the theorem assumptions, approximate quantitative agreement with the Born rule is nevertheless achieved. In the following sections, it will be shown that this allows us to reproduce many interesting phenomena that are otherwise thought to have no classical interpretation.

## 4 Unitary Transformations

A quantum state $|\psi\rangle$ is transformed to the state $U|\psi\rangle$ via a unitary operator $U$ representing the dynamics of the system, say, or an act of measurement in a particular basis. Representing the state by the complex amplitude vector $\boldsymbol{a}$, we may perform a similar transform to the vector $U\boldsymbol{a}$. The question at hand now is whether $U\boldsymbol{a}$ is a faithful statistical representation of $U|\psi\rangle$. To begin to answer that question, we consider the following.

**Lemma 1** *If $\mathbf{w} = \sigma\mathbf{z}$ is a complex Gaussian random vector with mean $\mathbf{0}$ and covariance $\sigma^2 I$ and $U$ is a unitary matrix, then $U\mathbf{w}$ has the same distribution as $\mathbf{w}$.*

*Proof* Since $U\mathbf{w}$ is a linear transformation of $\mathbf{w}$, it is also a complex Gaussian random vector, defined uniquely by its mean and covariance. By linearity, the mean is $\mathrm{E}[U\mathbf{w}] = U\mathrm{E}[\mathbf{w}] = \mathbf{0}$ and the covariance is $\mathrm{E}[U\mathbf{w}(U\mathbf{w})^\dagger] = U\mathrm{E}[\mathbf{w}\mathbf{w}^\dagger]U^\dagger = \sigma^2 I$. □

**Corollary 1** *If $\mathbf{w} = \sigma\mathbf{z}/\|\mathbf{z}\|$ and $U$ is a unitary matrix, then $U\mathbf{w}$ has the same distribution as $\mathbf{w}$.*

*Proof* Note that $U\mathbf{w} = \sigma U\mathbf{z}/\|\mathbf{z}\| = \sigma U\mathbf{z}/\|U\mathbf{z}\|$. Since $U\mathbf{z}$ has the same distribution as $\mathbf{z}$, the same is also true of $U\mathbf{w}$ and $\mathbf{w}$. □

We are now ready to introduce the main result of this section, regarding the relationship between $U\mathbf{a}$ and $U|\psi\rangle$.

**Theorem 4** *Let $U$ be any unitary matrix. If $\mathbf{a} = s\boldsymbol{\alpha} + \mathbf{w}$ and either $\mathbf{w} = \sigma\mathbf{z}$ or $\mathbf{w} = \sigma\mathbf{z}/\|\mathbf{z}\|$, then the detection probabilities for $U\mathbf{a}$ are given by $P_n(U\boldsymbol{\alpha}, \gamma)$ for $n \in \{0, 1, \ldots, N, \infty\}$.*

*Proof* The result follows directly from Lemma 1, Corollary 1, and the linearity of $U$. □

An important consequence of Theorem 4 is that, if the Born rule holds for all $\boldsymbol{\alpha}$ in the standard basis, then it holds for measurements in any basis, since they are related solely by a unitary transformation. In a more restrictive sense, if the Born rule holds for only a subset of all possible states and the unitary transform used to produce that particular measurement keeps the state within that subset, then the Born rule still applies for the new measurement basis.

To perform a measurement of an observable represented by a matrix $A$, we identify an associated unitary matrix $U$ such that $U^\dagger A U = \Lambda = \mathrm{diag}([\lambda_1, \ldots, \lambda_N])$ is diagonal. Let $A : \mathbb{C}^N \to \mathbb{R}$ be an associated random variable (i.e., measurable function) on the Borel subsets of $\mathbb{C}^N$ such that, given a complex amplitude vector $\mathbf{a}$, the outcome of the measurement is $A(\mathbf{a})$, which we define as follows [21]. Given $\mathbf{a} \in \mathbb{C}^N$, if $|(U^\dagger\mathbf{a})_n| > \gamma$ and $|(U^\dagger\mathbf{a})_{n'}| \leq \gamma$ for all $n' \neq n$, then we say that $A(\mathbf{a}) = \lambda_n$; otherwise, $A(\mathbf{a})$ is left undefined.

### 4.1 Example of Measurements in Orthogonal Bases

Using the example of Eq. (8), let us consider measurements of the Pauli spin operators $I$, $X$, $Y$, and $Z$, where

$$X = \begin{pmatrix} 0 & 1 \\ 1 & 0 \end{pmatrix}, \quad Y = \begin{pmatrix} 0 & -i \\ i & 0 \end{pmatrix}, \quad Z = \begin{pmatrix} 1 & 0 \\ 0 & -1 \end{pmatrix}. \tag{12}$$

A corresponding set of unitary matrices for diagonalizing $X$, $Y$, $Z$ are $H$, $V$, $I$, respectively, where

$$H = \frac{1}{\sqrt{2}} \begin{pmatrix} 1 & 1 \\ 1 & -1 \end{pmatrix}, \quad V = \frac{1}{\sqrt{2}} \begin{pmatrix} 1 & 1 \\ i & -i \end{pmatrix}. \tag{13}$$

Note that $H$ is the Hadamard matrix, representing the action of a beamsplitter. The matrix $V$ may be interpreted similarly, albeit with a different phase convention.

For definiteness, suppose $\gamma = \sigma = 1$, $s = \sqrt{2} - 1$, and $\boldsymbol{w}$ is drawn from $\boldsymbol{z}/\|\boldsymbol{z}\|$. Specifically, supposed $\boldsymbol{w} = [0.2197 - 0.7169i, -0.5290 + 0.3974i]^\mathsf{T}$ is a particular realization. This choice of values allows us to use Theorem 1, so that we are guaranteed that at most one threshold-crossing event occurs. To measure $Z$, say, we (trivially) apply $I$ to $\boldsymbol{a}$ and examine the component magnitudes. In this case, $|a_1| = 0.9570$ and $|a_2| = 0.6616$, so, as it turns out, there is no detection and, hence, no measurement outcome. In other words, $Z(\boldsymbol{a})$ is, in this case, undefined.

Now suppose instead that we have $\boldsymbol{w} = [0.5186 + 0.3818i, -0.6876 + 0.3354i]^\mathsf{T}$. In this case, $|a_1| = 1.0079$ and $|a_2| = 0.7650$, so there is a single detection indicating that $Z(\boldsymbol{a}) = +1$. (Indeed, this is the only possible outcome, given that there is a detection.) If a measurement of $X$ had been performed, we would have applied the unitary transformation $H^\dagger = H$ to $\boldsymbol{a}$ to obtain $H\boldsymbol{a} = [0.1734 + 0.5072i, 1.1458 + 0.0328i]$, so $|(H\boldsymbol{a})_1| = 0.5360$, $|(H\boldsymbol{a})_2| = 1.1463$, and the outcome $X(\boldsymbol{a}) = -1$ would have been obtained. To measure $Y$, we would apply $V^\dagger$ to $\boldsymbol{a}$ and obtain $V^\dagger \boldsymbol{a} = [0.8968 + 0.7561i, 0.4224 - 0.2162i]^\mathsf{T}$. Thus, $|(V^\dagger \boldsymbol{a})_1| = 1.1730$ and $|(V^\dagger \boldsymbol{a})_2| = 0.4745$, so $Y(\boldsymbol{a}) = +1$. In each case the measurement outcome is uniquely and counterfactually determined by $\boldsymbol{a}$.

### 4.2 Quantum State Inference

Now consider computing the expectation values of $X$, $Y$, and $Z$, conditioned on a single detection, when $\boldsymbol{\alpha}$ is one of $[1, 0]^\mathsf{T}$, $[0, 1]^\mathsf{T}$, or $\frac{1}{\sqrt{2}}[1, 1]^\mathsf{T}$. Note that the application of the corresponding unitary transformations $H$, $V^\dagger$, and $I$ will transform $\boldsymbol{\alpha}$ into a vector such that, again, each component is either zero or of the same magnitude. For example, if $\boldsymbol{\alpha} = [0, 1]^\mathsf{T}$, then $V^\dagger \boldsymbol{\alpha} = \frac{1}{\sqrt{2}}[-i, +i]^\mathsf{T}$. Provided that $\boldsymbol{w} = \sigma \boldsymbol{z}/\|\boldsymbol{z}\|$ and $\sigma \leq \gamma < \sqrt{s^2/2 + \sigma^2}$, the results of Theorem 3 will then hold and the conditional probabilities will match those of the Born rule. Consequently, the observed expectation values will be $\mathrm{E}[X] = \langle \psi | X | \psi \rangle$, $\mathrm{E}[Y] = \langle \psi | Y | \psi \rangle$, and $\mathrm{E}[Z] = \langle \psi | Z | \psi \rangle$.

Now, any two-dimensional operator can be written as a linear combination of $I$, $X$, $Y$, and $Z$. In particular, the quantum state operator $\rho$ may be written as

$$\rho = \frac{I + \mathrm{Tr}(\rho X) X + \mathrm{Tr}(\rho Y) Y + \mathrm{Tr}(\rho Z) Z}{2}. \tag{14}$$

Any quantum state, pure or mixed, may be written in this manner. For a pure state $\rho = |\psi\rangle \langle \psi|$, so $\mathrm{Tr}(\rho A) = \langle \psi | A | \psi \rangle$. Statistically, this corresponds to the expectation value E[$A$]. Let us therefore define the inferred quantum state operator

$$\tilde{\rho} = \frac{I + \mathrm{E}[X]X + \mathrm{E}[Y]Y + \mathrm{E}[Z]Z}{2}. \tag{15}$$

By using sample means to estimate the expectation values, the above expression allows a method for empirically deducing the quantum state from measurement, a process known as quantum state tomography (QST) [22,23]. As might be inferred from QST, then, the classical random vector $\boldsymbol{a}$ is statistically equivalent to the quantum state $|\psi\rangle$, since $\tilde{\rho} = \rho$ for these three choices of $\boldsymbol{\alpha}$.

QST is not the whole story, though. It has been noted that QST is inadequate to uniquely identify the underlying quantum state [24], and this example illustrates that fact. Consider a measurement of the operators $B_\pm = \mp(X \pm Z)/\sqrt{2}$, which are diagonalized by

$$W_\pm = \begin{pmatrix} \frac{\sqrt{2}\pm 1}{\sqrt{4\pm\sqrt{8}}} & -\frac{\sqrt{2}\mp 1}{\sqrt{4\mp\sqrt{8}}} \\ \frac{1}{\sqrt{4\pm\sqrt{8}}} & \frac{1}{\sqrt{4\mp\sqrt{8}}} \end{pmatrix}, \tag{16}$$

so that $W_\pm^\dagger B_\pm W_\pm = \mathrm{diag}([-1, +1])$. In this case, we find that for $\boldsymbol{\alpha} = [1, 0]^\mathsf{T}$, say, $p_1(\boldsymbol{\alpha}, \gamma) \approx 0.7048$ and $p_2(\boldsymbol{\alpha}, \gamma) \approx 0.2952$. So, E($B_\pm$) $\approx -0.4096$, but $\mathrm{Tr}(\rho B_\pm) = -1/\sqrt{2} \approx -0.7071$. Thus, even though the inferred quantum state is correct, the model does not predict exactly the right statistics for all observables. This example underscores the difficulty in verifying, empirically, that a given quantum state has, indeed, been correctly prepared.

### 4.3 Projective Subspace Measurements

If the observable to be measured is not resolvable into a nondegenerate eigenvector basis, then we must define its measurement more generally as a set of projections onto two or more subspaces within the larger Hilbert space. Let $\Pi_1, \ldots, \Pi_M$ be such a set of projections, where $M \leq N$ and $\Pi_1 + \cdots + \Pi_M = I$ is the identity. To perform a measurement, we project the vector $\boldsymbol{a}$, representing a particular realization, onto this set. A detection of projection $m$ is said to occur if $\|\Pi_m \boldsymbol{a}\| > \gamma$ while $\|\Pi_n \boldsymbol{a}\| \leq \gamma$ for all $n \neq m$. If $\Pi_m = |m\rangle\langle m|$, this reduces to the previous definition of measurement.

For example, suppose $\alpha = [0, 1, 1, 0]^\mathsf{T}/\sqrt{2}$ represents an entangled state. Let $\Pi_1 = |1\rangle\langle 1| + |2\rangle\langle 2|$ and $\Pi_2 = |3\rangle\langle 3| + |4\rangle\langle 4|$ be the two projections. We then find that $\|\Pi_1 \boldsymbol{a}\|^2 = |w_1|^2 + |(s/\sqrt{2}) + w_2|^2$ and $\|\Pi_2 \boldsymbol{a}\|^2 = |(s/\sqrt{2}) + w_3|^2 + |w_4|^2$. By symmetry, we see that the probability $\Pr[\|\Pi_1 \boldsymbol{a}\| > \gamma, \|\Pi_2 \boldsymbol{a}\| \leq \gamma]$ equals $\Pr[\|\Pi_2 \boldsymbol{a}\| > \gamma, \|\Pi_1 \boldsymbol{a}\| \leq \gamma]$; thus, a single detection is equally likely for either event. Conditioned on the actual occurrence of a single detection, we find that the two outcomes are perfectly anticorrelated, with each outcome having a probability of one-half.

Projective subspace measurements may be used to describe spacially separated measurements. Given $\Pi_1$, $\Pi_2$, and $\boldsymbol{a}$ as defined above, let $\Pi_1 \boldsymbol{a}$ be the portion associated with particle 1 and $\Pi_2 \boldsymbol{a}$ that of particle 2. We may measure an observable $X$, say, on particle 1 by applying $H$ to the projected state $\Pi_1 \boldsymbol{a}$ and observing if one of the two resulting component amplitudes exceeds the detection threshold. A similar procedure may be applied to particle 2, independent of particle 1. This approach will later be used in Sect. 6.2 to provide a classical analog to experimental tests of quantum nonlocality.

## 5 Quantum Contextuality

Quantum contextuality refers to the apparent dependence of measurement outcomes on what other, compatible, measurements one happens to choose to perform. It is closely connected to quantum nonlocality [25] and is believed to be important in the efficacy of certain quantum computing algorithms [26]. Recently, it has also been the subject of several experimental tests [27–29]. The concept is perhaps best understood in terms of the following example.

Suppose we have a set of nine operators, arranged in a square as follows:

$$\begin{array}{ccc} X \otimes I & I \otimes X & X \otimes X \\ I \otimes Y & Y \otimes I & Y \otimes Y \\ X \otimes Y & Y \otimes X & Z \otimes Z \end{array}$$

These operators constitute the famous Mermin–Peres "magic square" [30]. Using the fact that the Pauli matrices are involutions and that $XY = iZ$, it is readily verified that the product of the three operators in each row as well as that in the first two columns, is $I \otimes I$. For the third column, however, we note that $(X \otimes X)(Y \otimes Y)(Z \otimes Z) = -I \otimes I$.

According to the Kochen–Specker theorem, it is impossible to replace each of the nine observables with a definite value of either $+1$ or $-1$ (their two eigenvalues) in a consistent manner such that these same product relations hold. (This is readily proven or can be verified directly by simply trying all $2^9$ possible assignments.) From the perspective of quantum mechanics, this may seem odd: Since the three operators in each row and column are mutually commuting, they may be measured simultaneously. From the aforementioned product relations, the product of outcomes for each row measurement, and for the first two column measures, will always be $+1$, while that of the third column will always be $-1$. So, measurement reveals definite values that are consistent with the product relations, but no consistent assignment can be made across the square. Furthermore, this result holds independently of the prepared quantum state, or even of whether it is pure or mixed.

As pointed out in [31], the resolution of this paradox lies in the fact that, according to quantum mechanics, different probability measures apply to the six different choices of measurement bases. From a deterministic or hidden-variable perspective, one interpretation of this fact would be that the physical process of measurement induces a dynamical change in the hidden variable state such that the resulting post-measurement distribution is changed.

An alternate, and perhaps simpler, interpretation is possible if one considers measurement to be the threshold detection process considered here. In that case, the six post-measurement probability distributions are just the conditional probabilities, given that a single detection (for each observable) has occurred. It remains, then, to verify that this scheme does, in fact, work. To that end, it may be insightful to illustrate these properties explicitly via numerical simulations.

### 5.1 Monte Carlo Verification

A Monte Carlo scheme was devised to verify numerically that the detector model satisfies the properties of the magic square. For each Monte Carlo run, a random quantum state of the form $\boldsymbol{\alpha} = \boldsymbol{z}/\|\boldsymbol{z}\|$, with $N = 4$, was drawn. Next, a random realization of $M = 2^{20} = 1\,048\,576$ independent noise vectors of the form $\boldsymbol{w} = \sigma \boldsymbol{z}/\|\boldsymbol{z}\|$, with $\sigma = 1$, was drawn. From this, a set of $M$ complex amplitude vectors of the form $\boldsymbol{a} = s\boldsymbol{\alpha} + \boldsymbol{w}$, with $s = (\sqrt{2}-1)\sigma$, was created.

For each of the $M$ realizations of $\boldsymbol{a}$, six measurements were performed, corresponding to the three rows and three columns of the magic square. For each of the six measurements, a common unitary matrix $U$ was constructed that diagonalizes all three observables. The quantity $\boldsymbol{a}' = U^\dagger \boldsymbol{a}$ was then computed. For example, in the case of Row 1, it suffices to use $U_{R1} = H \otimes H$, while for Row 2 $U_{R2} = V \otimes V$ diagonalizes the observables. For Columns 1 and 2 we may use $U_{C1} = H \otimes V$ and $U_{C2} = V \otimes H$, respectively.

In the case of Column 3, where the three observables are $X \otimes X$, $Y \otimes Y$, and $Z \otimes Z$, the construction of $U_{C3}$ is not as straightforward. Each is diagonalized by $U = H \otimes H$, $V \otimes V$, and $I \otimes I$, respectively; however, none of these will diagonalize all three. Rearranging the columns of $U$ to produce $U' = [U(:, 4)\ U(:, 1)\ U(:, 2)\ U(:, 3)]$, where $U(:, n)$ is the $n^{\text{th}}$ column, produces an alternate unitary matrix which also diagonalizes $X \otimes X$. Finally, applying a second transformation yields $U_{C3} = U'(I \otimes H)$, which diagonalizes with respect to $Y \otimes Y$ and $Z \otimes Z$ as well. A similar procedure is required for Row 3.

In summary, the following unitary matrices were used for the rows

$$U_{R1} = H \otimes H \tag{17a}$$
$$U_{R2} = V \otimes V \tag{17b}$$
$$U_{R3} = \frac{1}{\sqrt{2}} \begin{pmatrix} 1 & 0 & 1 & 0 \\ 0 & -i & 0 & -i \\ 0 & -1 & 0 & 1 \\ i & 0 & -i & 0 \end{pmatrix} \tag{17c}$$

and columns

$$U_{C1} = H \otimes V \tag{18a}$$
$$U_{C2} = V \otimes H \tag{18b}$$
$$U_{C3} = \frac{1}{\sqrt{2}} \begin{pmatrix} 1 & 0 & 1 & 0 \\ 0 & -1 & 0 & -1 \\ 0 & -1 & 0 & 1 \\ 1 & 0 & -1 & 0 \end{pmatrix} \tag{18c}$$

If exactly one component of $\boldsymbol{a}'$ fell above the threshold $\gamma = \sigma$, then a detection was deemed to have occurred and the observable was assigned the value of the corresponding eigenvalue (either $+1$ or $-1$). If there was no detection, then no measurement was reported. By Theorem 3, we do not expect more than one component of $\boldsymbol{a}'$ to fall above the threshold. Thus, for each of the six sets of observables, there were $K \leq M$ reported triple values $[g_1(m_k), g_2(m_k), g_3(m_k)]$, for $k = 1, \ldots, K$, and an index (of length $K$) of which of the $M$ realizations resulted in a detection. These indices will be denoted $R_1, R_2, R_3$ for the three rows and $C_1, C_2, C_3$ for the three columns.

According to the Kochen–Specker theorem, we expect

$$(R_1 \cap R_2 \cap R_3) \cap (C_1 \cap C_2 \cap C_3) = \varnothing . \tag{19}$$

Furthermore, we expect that, say, for all $m \in C_3$ (i.e., for all detections when Column 3 is measured)

$$g_1(m) \, g_2(m) \, g_3(m) = -1 , \tag{20}$$

while for all $m \in R_1$

$$g_1(m) \, g_2(m) \, g_3(m) = +1 . \tag{21}$$

We expect the latter result to hold for $R_2, R_3, C_1$, and $C_2$ as well. A total of $2^{16} = 65\,536$ Monte Carlo runs were performed, each with a different random quantum state, and the above properties were examined. In all, cases, the required conditions were satisfied exactly.

As a specific example, consider the realization $\boldsymbol{a} = [-0.3151+0.5498i, -0.9092+0.1208i, -0.0581 - 0.5120i, 0.4560 - 0.3460i]^\mathsf{T}$. For this case, a measurement of Row 1 yields $(-1, 1, -1)$, Row 2 $(1, 1, 1)$, Row 3 $(-1, 1, -1)$, Column 1 $(-1, 1, -1)$, and Column 2 $(1, 1, 1)$. For Column 3, however, there are no detections, as $\boldsymbol{a}' = U_{C3}^\dagger \boldsymbol{a} = [0.0996+0.1441i, 0.6840+0.2766i, -0.5453+0.6334i, 0.6018-0.4475i]^\mathsf{T}$, and, hence, no measurement outcome. So, $\boldsymbol{a}$ is contained in all index sets except $C_3$, in accordance with the Kochen–Specker theorem.

## 5.2 Understanding the Magic Square

Let us consider these results more deeply. Each of the six unitary transformations diagonalizes the three observables in the corresponding row or column. For Row 3, this means

$$U_{R3}^\dagger (X \otimes Y) U_{R3} = \mathrm{diag}([+1, +1, -1, -1]) \tag{22a}$$

$$U_{R3}^\dagger (Y \otimes X) U_{R3} = \mathrm{diag}([+1, -1, -1, +1]) \tag{22b}$$

$$U_{R3}^\dagger (Z \otimes Z) U_{R3} = \mathrm{diag}([+1, -1, +1, -1]) \tag{22c}$$

while, for Column 3, we have

$$U_{C3}^\dagger (X \otimes X) U_{C3} = \text{diag}([+1, +1, -1, -1]) \tag{23a}$$

$$U_{C3}^\dagger (Y \otimes Y) U_{C3} = \text{diag}([-1, +1, +1, -1]) \tag{23b}$$

$$U_{C3}^\dagger (Z \otimes Z) U_{C3} = \text{diag}([+1, -1, +1, -1]) \tag{23c}$$

Suppose we measure Row 3 and obtain an outcome of $+1$ for $Z \otimes Z$. Since measurement outcomes are based on amplitude threshold crossings, and we have conditioned on single-detection events, we know that, for $\boldsymbol{a}' = U_{R3}^\dagger \boldsymbol{a}$, either $|a_1'| > \gamma$ or $|a_3'| > \gamma$. The outcomes for the other two observables, $X \otimes Y$ and $Y \otimes X$, could then be either $+1, +1$ (if $|a_1'| > \gamma$) or $-1, -1$ (if $|a_3'| > \gamma$), respectively. The outcome of, say, $X \otimes Y$ therefore uniquely determines the outcome of the other, and the product of all three is thereby guaranteed to be $+1$. Had we instead measured $-1$ for $Z \otimes Z$, a similar outcome would have been obtained.

Now suppose, using the same value of $\boldsymbol{a}$, we had measured Column 3 instead. It is possible that there are no detections, but we may suppose that there is one. Let us suppose further that the outcome of measuring $Z \otimes Z$ is $+1$ as well. This means that, for $\boldsymbol{a}'' = U_{C3}^\dagger \boldsymbol{a}$, either $|a_1''| > \gamma$ or $|a_3''| > \gamma$. For the other two observables, $X \otimes X$ and $Y \otimes Y$, the possible outcomes are $+1, -1$ (if $|a_1''| > \gamma$) and $-1, +1$ (if $|a_3''| > \gamma$), respectively. Again, the outcome of one observable, either $X \otimes X$ or $Y \otimes Y$, uniquely determines the outcome of the other, and, in either case, the product of all three is $-1$. Measuring $-1$ for $Z \otimes Z$ would, of course, have led to a similar outcome.

## 6 Entanglement

A quantum state $|\psi\rangle \in \mathcal{H} = \mathcal{H}_1 \otimes \mathcal{H}_2$, where $\mathcal{H}_1$ and $\mathcal{H}_2$ are Hilbert spaces and $\mathcal{H}$ is their tensor product space, is said to be *entangled* (with respect to $\mathcal{H}_1$ and $\mathcal{H}_2$) if there do not exist substates $|\psi_1\rangle \in \mathcal{H}_1$ and $|\psi_2\rangle \in \mathcal{H}_2$ such that $|\psi\rangle = |\psi_1\rangle \otimes |\psi_2\rangle$. A state which is not entangled is said to be *separable*.

This is a mathematical definition concerning properties of vector spaces which, taken at face value, suggests that there are plenty of *classical* systems that are entangled. For example, the acoustic field of a general, coupled mode solution to sound propagation in the ocean is, in this sense, an entangled state (though perhaps *nonseparable* would be a more accurate description). The term *classical entanglement* has been suggested to describe certain (classical) coherent optical states that exhibit properties similar to that of entangled quantum systems [32–34], though the lack of actual statistical outcomes makes this association somewhat dubious. Classical entanglement, as it is construed in these works, refers only to nonseparability. In the context of quantum mechanics, and in keeping with its original use by E. Schrödinger, entanglement connotes statistical correlations [35].

In what follows, we will consider *statistical* manifestations of entanglement arising from the proposed detector-based model. Given an entangled (i.e., non-separable) design state $|\psi\rangle$ and corresponding random vector $\boldsymbol{a}$, we would like to know whether the latter is entangled, and in what sense. Rather than appeal to vector space properties, we look instead to an equivalence of statistical predictions. In particular, we will examine the statistics of "detector clicks" for various observables.

A bit of notation may help. Let $|\uparrow\rangle = [1, 0]^T$ and $|\downarrow\rangle = [0, 1]^T$ denote the eigenvectors of $Z$, with eigenvalues $+1$ and $-1$, respectively. These may be interpreted as orthogonal polarizations of a single photon. The states $|\uparrow\rangle$ and $|\downarrow\rangle$ may also be viewed as the qubit states $|0\rangle$ and $|1\rangle$, respectively, in the computational basis [36]. If $\mathcal{H}_1 = \mathcal{H}_2 = \text{span}\{|\uparrow\rangle, |\downarrow\rangle\}$, then $\mathcal{H} = \mathcal{H}_1 \otimes \mathcal{H}_2$ is spanned by the following (separable) orthonormal basis states

$$|1\rangle = |\uparrow\rangle \otimes |\uparrow\rangle = |\uparrow\uparrow\rangle, \quad |2\rangle = |\uparrow\rangle \otimes |\downarrow\rangle = |\uparrow\downarrow\rangle$$
$$|3\rangle = |\downarrow\rangle \otimes |\uparrow\rangle = |\downarrow\uparrow\rangle, \quad |4\rangle = |\downarrow\rangle \otimes |\downarrow\rangle = |\downarrow\downarrow\rangle$$

This four-dimensional (i.e., two-qubit) space is the one we will investigate.

### 6.1 Bell States

The Bell states are a set of four maximally entangled states that also form an orthonormal basis for our four-dimensional Hilbert space, $\mathcal{H}$. They are given by

$$|\phi_1\rangle = \frac{1}{\sqrt{2}}[|\uparrow\uparrow\rangle + |\downarrow\downarrow\rangle], \quad |\phi_2\rangle = \frac{1}{\sqrt{2}}[|\uparrow\downarrow\rangle + |\downarrow\uparrow\rangle]$$
$$|\phi_3\rangle = \frac{1}{\sqrt{2}}[|\uparrow\uparrow\rangle - |\downarrow\downarrow\rangle], \quad |\phi_4\rangle = \frac{1}{\sqrt{2}}[|\uparrow\downarrow\rangle - |\downarrow\uparrow\rangle]$$

Suppose $|\psi\rangle = |\phi_2\rangle$, so the component vector in the standard basis is $\boldsymbol{\alpha} = [0, 1, 1, 0]^T/\sqrt{2}$. Now consider the 16 Hermitian matrices composed of pair-wise tensor products of the four Pauli matrices $I$, $X$, $Y$, and $Z$; i.e., $I \otimes I$, $I \otimes X$, ..., $Z \otimes Z$. It is readily verified that the Hilbert-Schmidt inner product between any two different pairs is zero, as this property holds for the Pauli matrices themselves. Between themselves, they take on the value 4. Hence these operators, when suitably normalized, form a tomographically complete quorum set, which may be used to deduce the quantum state from their expectation values using QST.

Going further, one finds that the application of each member of the quorum on $\boldsymbol{\alpha}$ results in a vector that, like $\boldsymbol{\alpha}$, has components that are either zero or of equal magnitude. With suitable choices of $s$, $\gamma$, and $\boldsymbol{w}$, then, Theorem 1 or 3 holds, depending upon the choice of distribution for $\boldsymbol{w}$, and the distribution of detected outcomes follows the Born rule. Consequently, the expectation values of the random variables corresponding to each observable match the quantum predictions exactly (in the case of Theorem 3) or asymptotically (in the case of Theorem 1). Tomographically, then, the random vector $\boldsymbol{a} = s\boldsymbol{\alpha} + \boldsymbol{w}$ is equivalent to the entangled state $|\psi\rangle$. It is straightforward to show that a similar result holds for the other three Bell states.

For example, let $s = (\sqrt{2} - 1)\sigma$, $\gamma = \sigma > 0$, and $\boldsymbol{w} = \sigma \boldsymbol{z}/\|\boldsymbol{z}\|$. Now suppose we measure $Z$ on both the first (left) particle and second (right) particle.. For the first particle, we will obtain either a projection onto the subspace $\text{Span}\{|\uparrow\uparrow\rangle, |\uparrow\downarrow\rangle\}$, if the result is $+1$, or $\text{Span}\{|\downarrow\uparrow\rangle, |\downarrow\downarrow\rangle\}$, if the result is $-1$. On the other hand, for the second

particle we will obtain a projection onto either the subspace Span$\{|\uparrow\uparrow\rangle, |\downarrow\uparrow\rangle\}$, if one measures $+1$, or Span$\{|\uparrow\downarrow\rangle, |\downarrow\downarrow\rangle\}$, if one measures $-1$. Of course, it is also possible that neither measurement yields a detection, but let us suppose that they both do. Since the random amplitudes of outcomes $|\uparrow\uparrow\rangle$ and $|\downarrow\downarrow\rangle$ are $|a_1| \leq \sigma$ and $|a_4| \leq \sigma$, respectively, these outcomes will never occur. Furthermore, since $E[|a_2|] = E[|a_3|]$, the outcomes $|\uparrow\downarrow\rangle$ and $|\downarrow\uparrow\rangle$ will be equally likely. Since, by Theorem 3, no more than one detection is possible, we conclude that, if we obtain $|\uparrow\rangle$ (i.e., $+1$) for particle 1, then we must obtain $|\downarrow\rangle$ (i.e., $-1$) for particle 2, and vice versa. The detected outcomes are thus perfectly anti-correlated, as one might expect for an entangled pair.

Of course, this doesn't work for every observable. Consider the Hermitian matrix $I \otimes B_+$, which is diagonalized by the unitary matrix $I \otimes W_+$. For $|\psi\rangle = |\phi_2\rangle$, the quantum probabilities $|\langle n|(I \otimes W_+)|\psi\rangle|^2$, for $n = 1, \ldots, 4$, are approximately 0.0732, 0.4268, 0.4268, 0.0732. The conditional detection probabilities are found numerically to be approximately 0.0364, 0.4608, 0.4641, 0.0388, with an uncertainty of about 0.004. Though similar, these values are only an approximation to the quantum predictions.

### 6.2 Violations of Bell's Inequality

Another hallmark of entanglement is the possibility of violating Bell's inequality. More precisely, the CHSH inequality is given by [37]

$$S_B = \left|E[AB] + E[AB']\right| + \left|E[A'B] - E[A'B']\right| \leq 2 , \tag{24}$$

where $A, A', B, B'$ are random variables bounded by unity and $E[\,\cdot\,]$ is the expectation with respect to some probability measure $P$. It is important to note that, in order for this inequality to hold, the same probability measure $P$ is used for all four expectation values. Thus,

$$E[AB] = \int A(\boldsymbol{a})B(\boldsymbol{a})dP(\boldsymbol{a}) \tag{25a}$$

$$E[AB'] = \int A(\boldsymbol{a})B'(\boldsymbol{a})dP(\boldsymbol{a}) \tag{25b}$$

$$E[A'B] = \int A'(\boldsymbol{a})B(\boldsymbol{a})dP(\boldsymbol{a}) \tag{25c}$$

$$E[A'B'] = \int A'(\boldsymbol{a})B'(\boldsymbol{a})dP(\boldsymbol{a}) \tag{25d}$$

The analogous expressions for quantum mechanics replace $E[AB]$, say, with $\langle AB \rangle = \langle \psi|AB|\psi\rangle$ for a fixed quantum state $|\psi\rangle$. In particular, if $|\psi\rangle$ is the Bell state $|\phi_2\rangle$, and the four observables are

$$A = Z \otimes I, \quad B = -I \otimes (X+Z)/\sqrt{2}$$
$$A' = X \otimes I, \quad B' = +I \otimes (X-Z)/\sqrt{2}$$

then
$$S_Q = |\langle AB\rangle + \langle AB'\rangle| + |\langle A'B\rangle - \langle A'B'\rangle| = 2\sqrt{2}\ . \qquad (26)$$

How can we reconcile this result with the CHSH inequality? As we have noted, the theorem applies to expectations that are with respect to the *same* probability measure. If the probability measures differ for each pair of observables, then the inequality need no longer hold. Now, in the model proposed here, expectations are with respect to *conditional* probability distributions, conditioned, that is, on single detections. Let these be denoted $E_1[AB]$, $E_2[AB']$, $E_3[A'B]$, and $E_4[A'B']$. This means that Bell's theorem does not apply and violations of the inequality are possible. It remains to ask whether
$$S_D = \big|E_1[AB] + E_2[AB']\big| + \big|E_3[A'B] - E_4[A'B']\big| \qquad (27)$$

can ever be greater than 2.

A numerical study was performed to investigate this possibility. As before, a random realization of $M = 2^{20}$ independent noise vectors of the form $\mathbf{w} = \sigma \mathbf{z}/\|\mathbf{z}\|$, with $\sigma = 1$, was drawn. Using $\boldsymbol{\alpha} = [0, 1, 1, 0]^T$, a set of $M$ complex-valued vectors of the form $\mathbf{a} = s\boldsymbol{\alpha} + \mathbf{w}$, with $s = (\sqrt{2} - 1)\sigma$, was created. A detection threshold of $\gamma = \sigma$ was used.

To this set the Hermitian conjugates of the unitary matrices $U_1 = I \otimes W_+$, $U_2 = I \otimes W_-$, $U_3 = H \otimes W_+$, and $U_4 = H \otimes W_-$ were applied separately to $\mathbf{a}$ for the observables $AB$, $AB'$, $A'B$, and $A'B'$, respectively. For each of the four measurements, the diagonalized matrix of eigenvalues is

$$U_1^\dagger(AB)U_1 = \mathrm{diag}([-1, +1, +1, -1]) \qquad (28a)$$
$$U_2^\dagger(AB')U_2 = \mathrm{diag}([+1, -1, -1, +1]) \qquad (28b)$$
$$U_3^\dagger(A'B)U_3 = \mathrm{diag}([-1, +1, +1, -1]) \qquad (28c)$$
$$U_4^\dagger(A'B')U_4 = \mathrm{diag}([+1, -1, -1, +1]) \qquad (28d)$$

Of the $M$ realizations of $\mathbf{a}$, typically only about 5 % resulted in a detection, though it was a different 5 % for each of the four observables. (This should not be confused with the detector efficiency, which is a measure of coincidence rates and no meaning in this context.) Let $I_1$ denote the set of values of $\mathbf{a}$, a subset of all $M$ realizations, that resulted in a detection for the observable $AB$. Define $I_2$, $I_3$, and $I_4$ similarly for $AB'$, $A'B$, and $A'B'$, respectively. Furthermore, let $I_{ij}$ denote the subset of $I_i$ for which the $j^{\text{th}}$ component exceeded the threshold. Note that $I_{i1}$, $I_{i2}$, $I_{i3}$, and $I_{i4}$ are mutually exclusive, and their union is $I_i$. Finally, let $n_{ij}$ denote the cardinality of $I_{ij}$ and $n_i$ the cardinality of $I_i$.

The results of the numerical simulation are summarized in Table 1. From these results we deduce mean values of $E_1[AB] = 0.8497$, $E_2[AB'] = 0.8440$, $E_3[A'B] = -0.8486$, and $E_4[A'B'] = 0.8481$, each with an uncertainty of about 0.004, computed as $1/\sqrt{n_i}$. Combining these results, we find

$$S_D = 3.39 \pm 0.016\ , \qquad (29)$$

**Table 1** Results of a numerical simulation to test violations of Bell's inequality using $\boldsymbol{w} = \sigma \boldsymbol{z}/\|\boldsymbol{z}\|$

| Observable | $n_{i1}$ | $n_{i2}$ | $n_{i3}$ | $n_{i4}$ | $n_i$ | Mean |
|---|---|---|---|---|---|---|
| $AB$ | 2,166 | 27,439 | 27,633 | 2,308 | 59,546 | 0.8497 |
| $AB'$ | 27,311 | 2,358 | 2,272 | 27,435 | 59,376 | 0.8440 |
| $A'B$ | 27,236 | 2,239 | 2,232 | 27,357 | 59,064 | −0.8486 |
| $A'B'$ | 27,415 | 2,241 | 2,252 | 27,252 | 59,160 | 0.8481 |

About 5 % of the $M = 2^{20}$ realizations resulted in a single detection where the uncertainties of the four correlations were added to arrive at the final uncertainty in $S_D$. This is clearly greater than 2 and, in fact, greater than the Tsirelson bound of $2\sqrt{2} = 2.8284$, which is an upper bound on quantum violations of the CHSH inequality.

**Table 2** Results of a numerical simulation to test violations of Bell's inequality using $\boldsymbol{w} = \sigma \boldsymbol{z}$. About 0.25 % of the $M = 2^{20}$ realizations resulted in a single detection

| Observable | $n_{i1}$ | $n_{i2}$ | $n_{i3}$ | $n_{i4}$ | $n_i$ | Mean |
|---|---|---|---|---|---|---|
| $AB$ | 250 | 1,075 | 1,091 | 225 | 2,641 | 0.6403 |
| $AB'$ | 1,062 | 246 | 216 | 1,104 | 2,628 | 0.6484 |
| $A'B$ | 1,036 | 228 | 193 | 1,091 | 2,548 | −0.6695 |
| $A'B'$ | 1,079 | 202 | 222 | 1,093 | 2,596 | 0.6733 |

A similar numerical study was performed using $\boldsymbol{w} = \sigma \boldsymbol{z}$, with $\sigma = 1$, $s = \sigma$, and $\gamma = 3\sigma$. The results of this study are summarized in Table 2, from which we deduce the following mean values: $E_1[AB] = 0.6403$, $E_2[AB'] = 0.6484$, $E_3[A'B] = -0.6695$, and $E_4[A'B'] = 0.6733$, each with an uncertainty of about 0.02. Combining these results, we find

$$S_D = 2.63 \pm 0.08 . \tag{30}$$

Although the correlations are not as strong, we do find again that Bell's inequality is violated.

These results demonstrate that a (classical) deterministic model can violate Bell's inequality. Such violations are made possible by the fact that the model is contextual, and this contextuality is, in turn, a consequence of our conditioning on single-detection events. In some cases, these violations can be larger than those predicted by quantum mechanics. This is so despite the fact that the Born rule is not perfectly reproduced for all observables concerned.

### 6.3 Local Realism

The notion that quantum mechanics is at odds with local realism first arose in the context of the Einstein–Podolsky–Rosen (EPR) paradox [38] and later by Bohm in terms of discrete states [39]. This paradox was recast by Bell [4] into an inequality that, he concluded, no local realistic theory could violate. A variation of this inequality was first tested by Clauser[40] and, in a later landmark experiment, by Aspect [41], with results in agreement with quantum predictions. This is generally regarded as conclusive evidence that quantum mechanics, and hence nature, is fundamentally nonlocal, despite the fact that it has been known for some time that violations of Bell's inequality are possible under the so-called detection loophole [42–44]. The analysis of the previous section reconfirms this through a specific example, but it did not address *local* realism, as the two observables were effectively measured as one. Although recent experiments, with detector efficiencies over 70 %, claim to have closed the detection loophole for photons, these results do not address violations of Bell's inequality but, rather, a little-known inequality due to Eberhard used to test local realism while accounting for detector inefficiencies [45–47].

Let us, then, consider an alternate scheme whereby the two observables are measured separately and independently. This corresponds to the usual sense of local realism in the context of Bell's inequality, namely, that the choice of $A$ or $A'$ and its outcome are not influenced by (nor do they influence) the choice of $B$ or $B'$ and its outcome. As is common in such discussions, we will describe this situation in terms of two familiar actors.

Suppose Alice and Bob, who live in different cities, both receive a letter upon which is written a particular realization, say, $s\boldsymbol{\alpha} + \boldsymbol{w} = [-0.165 + 0.2046i, 0.8316 + 0.6696i, 0.5690 - 0.2230i, 0.2321 - 0.1111i]^\mathsf{T}$, of the Bell state $\boldsymbol{\alpha} = [0, 1, 1, 0]^\mathsf{T}/\sqrt{2}$ with $s = (\sqrt{2} - 1)\sigma$, $\boldsymbol{w} = \sigma \boldsymbol{z}/\|\boldsymbol{z}\|$, and $\sigma = 1$. At a previously agreed upon time they each select a measurement to perform on it. Alice chooses either $A = Z \otimes I$ or $A' = X \otimes I$, while Bob chooses either $B = -I \otimes (X+Z)/\sqrt{2}$ or $B' = I \otimes (X-Z)/\sqrt{2}$. The choice and outcome are written down but not shared until later.

These measurements are performed, not by a device, but by simple pencil-and-paper calculations whereby Alice multiplies $s\boldsymbol{\alpha} + \boldsymbol{w}$ by either $(I \otimes I)^\dagger$, to measure $A$, or $(H \otimes I)^\dagger$, to measure $A'$, thereby obtaining the result $\boldsymbol{a}$ or $\boldsymbol{a}'$, respectively. Using a detection threshold $\gamma = \sigma$, she determines the result of the measurement for, say, $A$ by noting if either $|a_1|^2 + |a_2|^2 > \gamma^2$, in which case she writes "+1", or $|a_3|^2 + |a_4|^2 > \gamma^2$, in which case she writes "−1". This is so because

$$(I \otimes I)^\dagger A (I \otimes I) = \mathrm{diag}([+1, +1, -1, -1]) . \qquad (31)$$

If neither is the case, or if both are true, she simply writes "NaN" (Not a Number). The same procedure is followed when measuring $A'$, since

$$(H \otimes I)^\dagger A' (H \otimes I) = \mathrm{diag}([+1, +1, -1, -1]) . \qquad (32)$$

Bob proceeds in much the same manner as Alice, multiplying $s\boldsymbol{\alpha} + \boldsymbol{w}$ by either $(I \otimes W_+)^\dagger$, to measure $B$, or $(I \otimes W_-)^\dagger$, to measure $B'$ and denoting the respective results $\boldsymbol{b}$ and $\boldsymbol{b}'$. To determine the outcome of measuring, say, $B$, Bob examines whether $|b_2|^2 + |b_4|^2 > \gamma^2$, in which case he writes "+1", or $|b_1|^2 + |b_3|^2 > \gamma^2$, in which case he writes "−1". This is so because

$$(I \otimes W_+)^\dagger B (I \otimes W_+) = \mathrm{diag}([-1, +1, -1, +1]) . \qquad (33)$$

To measure $B'$, Bob examines whether $|b'_1|^2 + |b'_3|^2 > \gamma^2$, in which case he writes "+1", or $|b'_2|^2 + |b'_4|^2 > \gamma^2$, in which case he writes "−1", since

$$(I \otimes W_-)^\dagger B'(I \otimes W_-) = \text{diag}([+1, -1, +1, -1]) . \tag{34}$$

Like Alice, if there are no threshold crossings, or two threshold crossings, he simply writes "NaN" for the outcome.

Suppose Alice chooses to measure $A$. In this case, $\boldsymbol{a} = s\boldsymbol{\alpha} + \boldsymbol{w}$ and she finds that $|a_1|^2 + |a_2|^2 = 1.209 > \gamma^2$ while $|a_3|^2 + |a_4|^2 = 0.4397 \leq \gamma^2$, so she writes down "+1." Suppose further that Bob independently chooses to measure $B$ and thereby computes $\boldsymbol{b} = [0.1658 + 0.4453i, 0.8314 + 0.5403i, 0.6145 - 0.2485i, -0.0033 - 0.0173i]^\mathsf{T}$. Since $|b_2|^2 + |b_4|^2 = 0.9836 \leq \gamma^2$ and $|b_1|^2 + |b_3|^2 = 0.6651 \leq \gamma^2$, he records "NaN" — no valid outcome was obtained. Had Bob chosen to measure $B'$ instead, he would have found $\boldsymbol{b}' = [0.7052 + 0.6969i, 0.4707 + 0.0672i, 0.4322 - 0.1880i, -0.4369 + 0.1635i]^\mathsf{T}$. In that case, since $|b'_2|^2 + |b'_4|^2 = 0.4436 <= \gamma^2$ and $|b'_1|^2 + |b'_3|^2 = 1.2051 > \gamma^2$, he would have recorded "+1" instead of "NaN" for the outcome.

Now suppose Alice and Bob play this game many, many times. For each instance of $s\boldsymbol{\alpha} + \boldsymbol{w}$ they record which measurement they performed and whether they obtained "+1," "−1," or "NaN" as an outcome. When the game is over, they compare notes. All items on the list in which either Alice or Bob recorded "NaN" are struck out. Next, the results are grouped into four categories, corresponding to the four measurement combinations. Finally, the correlation is computed for each group.

A numerical study was performed along these lines, with $M = 2^{20}$ realizations of $s\boldsymbol{\alpha} + \boldsymbol{w}$ for each of the four measurement choices. In each case, about 30 % of the $M$ realizations resulted in a single detection for either Alice or Bob, and about 10 % of the $M$ realizations resulted in a single coincidence detection. This corresponded to a detector efficiency of about $\eta = 0.33$ (the ratio of coincident to single detections), which is comparable to that of a good quantum optics experiment.

Table 3 summarizes the results of this study. For example, the number of times Alice obtained ↑ (+1) and Bob obtained ↓ (−1) when she measured $A$ and he measured $B$ was 12069. The total number of coincidences for this pair of observables was 118251, resulting in a mean correlation of $E_1[AB] = 0.5833$. Computing these four mean correlations allows us to compute $S_D$, which was found to be

$$S_D = 2.3405 \pm 0.004 . \tag{35}$$

This result is, of course, larger than 2 and, so, violates Bell's inequality. As before, this was made possible by the fact that not all measurements resulted in a single coincidence detection for both Alice and Bob. This example, however, shows that local measurements made within a fully deterministic (i.e., classical) model can still violate Bell's inequality.

Interestingly, the violation is not as large as was found in Sect. 6.2. It seems, therefore that separated measurements have weaker correlations than joint measurements. Furthermore, if one uses $\boldsymbol{w} = \boldsymbol{z}$ instead of $\boldsymbol{w} = \boldsymbol{z}/\|\boldsymbol{z}\|$, as was done above, no violation is observed. Introducing correlations in the initial noise term therefore seems to have

**Table 3** Results of a numerical simulation to test violations of Bell's inequality under the local measurements

| Alice | Bob | ↑ ↑ | ↑ ↓ | ↓ ↑ | ↓ ↓ | Total | Mean |
|---|---|---|---|---|---|---|---|
| $A$ | $B$ | 46,523 | 12,069 | 12,402 | 46,924 | 118,251 | 0.5833 |
| $A$ | $B'$ | 46,633 | 12,171 | 12,372 | 46,819 | 118,196 | 0.5830 |
| $A'$ | $B$ | 12,226 | 46,674 | 46,795 | 12,119 | 117,935 | −0.5861 |
| $A'$ | $B'$ | 46,622 | 12,129 | 12,091 | 46,738 | 117,580 | 0.5880 |

About 10 % of the $M = 2^{20}$ realizations resulted in a single coincidence detection the effect of strengthening the correlations in the measured outcomes. This suggests that a different choice of noise distribution could lead to a higher efficiency and an even larger violation.

## 7 Conclusion

This paper introduces a simple deterministic model of quantum systems and quantum measurement that is capable of reproducing many phenomena typically regarded as having no classical analogue. The model associates a given pure quantum state $|\psi\rangle$ with a complex random vector $\boldsymbol{a}$ that is composed of a scaled version of the state's complex components, $s\boldsymbol{\alpha}$, and an additive complex noise term $\boldsymbol{w}$. Although not addressed here, mixed states may be modeled similarly using an ensemble of pure states.

A measurement is taken to be a single-threshold-crossing event ($|a_n| > \gamma$, $|a_{n'}| \leq \gamma$ for all $n' \neq n$); all other events are ignored. Taking the noise to be either a vector of independent complex Gaussians ($\boldsymbol{w} = \sigma\boldsymbol{z}$) or its normalized counterpart ($\boldsymbol{w} = \sigma\boldsymbol{z}/\|\boldsymbol{z}\|$), it was shown that, for suitable choices of $s$ and $\gamma$ relative to $\sigma$, the Born rule is recovered for states that are such that the components are either zero or equal in magnitude.

Measurements in other bases are performed by applying the corresponding unitary transformation to the vector $\boldsymbol{a}$. Using these properties, one can use quantum state tomography to deduce the equivalent quantum state from the statistics of single-detection events. Partial measurements over a complete set of projection operators are defined similarly, with the amplitude of the projection onto each subspace being used for threshold detection. In the case that the projections are formed from a complete orthonormal basis, this reduces to the above prescription for full measurements.

This model has been shown to be capable of reproducing several aspects of quantum contextuality and entanglement, including local violations of Bell's inequality. Common among these varied phenomena is the dependence of the underlying statistics on different, noncommuting sets of observables. This dependence has previously been known to give rise to contextuality. Here, this contextuality arises from the fact that we have conditioned on single-threshold-crossing events for our definition of measurement.

Although this model does exhibit many of the more interesting features of quantum phenomena, it is by no means complete. The Born rule is reproduced exactly in only a limited set of quantum states and is elsewhere only an approximation. Bounded Gaussian noise models appear to work better that unbounded ones, although both are capable of reproducing quantum phenomena.

While something of a mathematical contrivance, it is hoped that this model may form the basis of a more physical theory of quantum measurement. Further extensions of this work would include improving the noise model and associating the random complex vector to a particular physical model of some classical system, such as a stochastic electromagnetic field varying over space and time. Similarly, the artifice of a threshold detector is but a crude representation of light-matter interactions, which should be modeled in greater detail for a full physical theory. Proceeding in this manner, one might hope that better agreement over a broader set of quantum phenomena may be achieved.

**Acknowledgments** This work was supported by an Internal Research and Development grant from Applied Research Laboratories, The University of Texas at Austin. The author would like to thank Dr. M. Thrasher for his review of the manuscript and many helpful suggestions. The author would also like to thank the reviewers for their comments and recommended references, both of which have contributed to the quality of the final manuscript.

## Appendix

**Theorem 5** *Suppose the components of $\mathbf{w} = \sigma \mathbf{z}$ are independent and identically distributed complex Gaussian random variables with zero mean and variance $\sigma^2 > 0$. Then $|s\alpha_m| < |s\alpha_k|$ implies*

$$\lim_{\gamma \to \infty} \frac{P_m(\boldsymbol{\alpha}, \gamma)}{P_k(\boldsymbol{\alpha}, \gamma)} = 0 \,. \tag{36}$$

*Proof* From the assumed distribution of $\mathbf{w}$, it follows that $|a_1/\sigma|^2, \ldots, |a_N/\sigma|^2$ are independent, non-central $\chi^2$-distributed random variables, each with two degrees of freedom and noncentrality parameter $\lambda_i = |s\alpha_i/\sigma|^2$ for $i = 1, \ldots, N$.

Let $F_i(\gamma) = \Pr[|a_i| \leq \gamma] = \Pr[|a_i/\sigma|^2 \leq (\gamma/\sigma)^2]$ denote the cumulative distribution function for the $i^{\text{th}}$ component and note that

$$F_i(\gamma) = 1 - Q_1(\sqrt{\lambda_i}, \gamma/\sigma) \,, \tag{37}$$

where $Q_1(a, b)$ is the Marcum Q-function [48],

$$Q_1(a, b) = \int_b^\infty x \, e^{-(x^2 + a^2)/2} I_0(ax) \, dx \,, \tag{38}$$

and $I_0$ is the modified Bessel function of order 0. Note that $Q_1(a, 0) = 1$ and $Q_1(a, b) < 1$ for $b > 0$.

In the special case $a = 0$, corresponding to $s\alpha_i = 0$, we note that

$$Q_1(0, b) = e^{-b^2/2} \,. \tag{39}$$

Thus, for $s\alpha_i = 0$,

$$F_i(\gamma) = 1 - e^{-\gamma^2/(2\sigma^2)} \,. \tag{40}$$

For $s\alpha_i \neq 0$ there is no closed-form solution for $F_i$. As we are interested in the large-$\gamma$ limit, however, we may consider the behavior of $Q_1(a, b)$ for large $b$. In this regime, and for $a \neq 0$, the modified Bessel function in the integrand may be approximated as [49]

$$I_0(ax) \approx \frac{e^{ax}}{\sqrt{2\pi a x}} \,. \tag{41}$$

In this approximation, the Marcum Q-function becomes

$$Q_1(a, b) \approx \frac{1}{\sqrt{2\pi a}} \int_b^\infty \sqrt{x} \, e^{-(x-a)^2/2} dx \,. \tag{42}$$

This still does not admit a closed-form solution, but it may be bounded. Define the lower and upper bounding functions

$$Q_1^L(a, b) := \frac{1}{\sqrt{2\pi a}} \int_b^\infty e^{-(x-a)^2/2} dx \tag{43}$$

$$Q_1^U(a, b) := \frac{1}{\sqrt{2\pi a}} \int_b^\infty x \, e^{-(x-a)^2/2} dx \tag{44}$$

and note that, for $b$ sufficiently large,

$$Q_1^L(a, b) \leq Q_1(a, b) \leq Q_1^U(a, b) \,. \tag{45}$$

These two integrals do admit closed-form solutions, which are of the form

$$Q_1^L(a, b) = \frac{1}{\sqrt{4a}} \operatorname{erfc}\left(\frac{b-a}{\sqrt{2}}\right) \tag{46}$$

$$Q_1^U(a, b) = \frac{e^{-(b-a)^2/2}}{\sqrt{2\pi a}} + \sqrt{\frac{a}{4}} \operatorname{erfc}\left(\frac{b-a}{\sqrt{2}}\right) \tag{47}$$

where erfc is the complementary error function. The function erfc itself has the following asymptotic form:

$$\operatorname{erfc}(x) \sim \frac{e^{-x^2}}{x\sqrt{\pi}} \quad \text{for large } x \,. \tag{48}$$

Thus,

$$Q_1^L(a, b) \approx \frac{e^{-(b-a)^2/2}}{\sqrt{2\pi a}(b-a)} \qquad (49)$$

$$Q_1^U(a, b) \approx \frac{e^{-(b-a)^2/2}}{\sqrt{2\pi a}} + \sqrt{\frac{a}{2\pi}} \frac{e^{-(b-a)^2/2}}{b-a} \qquad (50)$$

From this result, we can see that $Q_1(a, b) \sim e^{-(b-a)^2/2}$ for large $b$. Thus, for large $\gamma$,

$$F_i(\gamma) \sim 1 - \exp\left[-\left(\gamma/\sigma - \sqrt{\lambda_i}\right)^2/2\right]. \qquad (51)$$

Now consider $P_n(\boldsymbol{\alpha}, \gamma)$, the probability that $|a_n| > \gamma$ is the only threshold crossing. Since the random variables $a_1, \ldots, a_N$ are independent, it may be written in terms of $F_i(\gamma)$ as simply

$$P_n(\boldsymbol{\alpha}, \gamma) = [1 - F_n(\gamma)] \prod_{i \neq n} F_i(\gamma). \qquad (52)$$

Now suppose $|s\alpha_m| < |s\alpha_k|$ and consider the ratio $P_m(\gamma)/P_k(\gamma)$ for $\gamma > 0$. This will be given by

$$\frac{P_m(\boldsymbol{\alpha}, \gamma)}{P_k(\boldsymbol{\alpha}, \gamma)} = \frac{1 - F_m(\gamma)}{1 - F_k(\gamma)} \frac{F_k(\gamma)}{F_m(\gamma)}. \qquad (53)$$

For large $\gamma$, the ratio $F_k(\gamma)/F_m(\gamma)$ is approximately unity and may be ignored. The remaining factors may be written in terms of Eq. (51), so that

$$\frac{P_m(\boldsymbol{\alpha}, \gamma)}{P_k(\boldsymbol{\alpha}, \gamma)} \approx \frac{\exp\left[-\left(\gamma/\sigma - \sqrt{\lambda_m}\right)^2/2\right]}{\exp\left[-\left(\gamma/\sigma - \sqrt{\lambda_k}\right)^2/2\right]}$$
$$= \exp[-(\sqrt{\lambda_k} - \sqrt{\lambda_m})\gamma/\sigma] e^{\lambda_k - \lambda_m}, \qquad (54)$$

which tends to zero as $\gamma \to \infty$, since $\lambda_k = |s\alpha_k|^2 > |s\alpha_m|^2 = \lambda_m$. □

## References

1. Bell, J.S.: On the problem of hidden variables in quantum mechanics. Rev. Mod. Phys. **38**, 447–452 (1966)
2. Genovese, M.: Research on hidden variable theories: a review of recent progress. Phys. Rep. **413**, 319–396 (2005)
3. Kochen, S., Specker, E.P.: The problem of hidden variables in quantum mechanics. J. Math. Mech. **17**, 59–87 (1967)
4. Bell, J.S.: On the Einstein Podolsky Rosen paradox. Physics **1**, 195–200 (1964)
5. Dragoman, D., Dragoman, M.: Quantum-Classical Analogies. Springer, Berlin (2004)
6. Khrennikov, A.: Ubiquitous Quantum Structure: From Psychology to Finance. Springer, Berlin (2010)
7. Spekkens, R.W.: Evidence for the epistemic view of quantum states: a toy theory. Phys. Rev. A **75**, 032110 (2007)
8. Pusey, M.F., Barrett, J., Rudolph, T.: On the reality of the quantum state. Nat. Phys. **8**, 475–478 (2012)

9. de la Peña, L., Cetto, A.M.: The Quantum Dice: An Introduction to Stochastic Electrodynamics. Springer, Berlin (1995)
10. Marshall, T., Santos, E.: Stochastic optics: a reaffirmation of the wave nature of light. Found. Phys. **18**(2), 185–223 (1988)
11. Brida, G., Genovese, M., Gramegna, M., Novero, C., Predazzi, E.: A first test of Wigner function local realistic model. Phys. Lett. A **299**, 121–124 (2002)
12. Brida, G., Genovese, M., Gramegna, M., Novero, C.: Experimental limit on spontaneous parametric up conversion. J. Mod. Opt. **50**(11), 1757–1762 (2003)
13. Scully, M.: Quantum Optics, 1st edn. Cambridge University Press, Cambridge (1997)
14. Garola, C.: A proposal for embodying quantum mechanics in a noncontextual framework by reinterpreting quantum probablilites. Foundations of Probability and Physics-5. AIP, vol. 1101, pp. 42–50 (2008)
15. Skyrms, B.: Counterfactual definiteness and local causation. Philos. Sci. **49**, 43–50 (1982)
16. Khrennikov, A.: Quantum probabilities and violation of CHSH-inequality from classical random signals and threshold type detection scheme. Prog. Theor. Phys. **128**, 31–58 (July 2012)
17. Bell, J.: Speakable and Unspeakable in Quantum Mechanics. Cambridge University Press, Cambridge (1987)
18. Gammaitoni, L., Hanggi, P., Jung, P., Marchesoni, F.: Stochastic resonance. Rev. Mod. Phys. **70**(1), 223–287 (1998)
19. Marshall, T.W., Santos, E.: Stochastic optics: a local realist analysis of optical tests of the Bell inequalities. Phys. Rev. A **39**, 6271–6283 (1989)
20. Greenberger, D.M., Horne, M.A., Zeilinger, A.: Bell's Theorem, Quantum Theory, and Conceptions of the Universe, pp. 73–76. Kluwer Academics, Dordrecht (1989)
21. Dudley, R.M.: Real Analysis and Probability, 1st edn. Chapman & Hall, New York (1989)
22. Fano, U.: Description of states in quantum mechanics by density matrix and operator techniques. Rev. Mod. Phys. **29**, 74–93 (1957)
23. Ježek, M., Fiurášek, J., Hradil, Z.: Quantum inference of states and processes. Phys. Rev. A **68**, 012305 (2003)
24. Bužek, V., Drobný, G., Derka, R., Adam, G., Wiedemann, H.: Quantum state reconstruction from incomplete data. Chaos Solitons Fractals **10**, 981–1074 (1999)
25. Cabello, A.: Proposal for revealing quantum nonlocality via local contextuality. Phys. Rev. Lett. **104**, 220401 (2010)
26. Howard M., Wallman, J., Veitch, V., Emerson, J.: Contextuality supplies the magic for quantum computation. arXiv:1401.4174v1 (2014)
27. Huang, Y.-F., Li, C.-F., Zhang, Y.-S., Pan, J.-W., Guo, G.-C.: Experimental test of the Kochen–Specker theorem with single photons. Phys. Rev. Lett. **90**, 250401 (2003)
28. Hasegawa, Y., Loidl, R., Badurek, G., Baron, M., Rauch, H.: Quantum contextuality in a single-neutron optical experiment. Phys. Rev. Lett. **97**, 230401 (2006)
29. Kirchmair, G., Zähringer, F., Gerritsma, R., Kleinmann, M., Gühne, O., Cabello, A., Blatt, R., Roos, C.: State-independent experimental test of quantum contextuality. Nature **460**, 494–497 (2009)
30. Mermin, D.: Hidden variables and the two theorems of John Bell. Rev. Mod. Phys. **65**, 803–815 (1993)
31. La Cour, B.R.: Quantum contextuality in the Mermin–Peres square: a hidden variable perspective. Phys. Rev. A **79**, 012102 (2009)
32. Spreeuw, R.J.C.: A classical analogy of entanglement. Found. Phys. **28**(3), 361–374 (1998)
33. Chowdhury, P., Majumdar, A.S., Agarwal, G.S.: Nonlocal continuous-variable correlations and violation of Bell's inequality for light beams with topological singularities. Phys. Rev. A **88**, 013830 (2013)
34. Ghose, P., Mukherjee, A.: Novel states of classical light and noncontextuality. Advanced Science, Engineering and Medicine **6**(2), 246–251 (2014)
35. Horodecki, R., Horodecki, P., Horodecki, M., Horodecki, K.: Quantum entanglement. Rev. Mod. Phys. **81**(2), 865–942 (2009)
36. Nielsen, M.A., Chuang, I.L.: Quantum Computation and Quantum Information. Cambridge University Press, Cambridge (2000)
37. Clauser, J.F., Horne, M.A., Shimony, A., Holt, R.A.: Proposed experiment to test local hidden-variable theories. Phys. Rev. Lett. **23**, 880–884 (1969)
38. Einstein, A., Podolsky, B., Rosen, N.: Can the quantum-mechanical description of physical reality be considered complete? Phys. Rev. **47**, 777–780 (1935)

39. Bohm, D.: Quantum Theory. Prentice Hall, New York (1951)
40. Clauser, J.F., Horne, M.A.: Experimental consequences of objective local theories. Phys. Rev. D **10**, 526–535 (1974)
41. Aspect, A., Grangier, P., Roger, G.: Experimental realization of Einstein–Podolsky–Rosen–Bohm Gedanken experiment: a new violation of Bell's inequalities. Phys. Rev. Lett. **49**(2), 91–94 (1982)
42. Pearle, P.M.: Hidden-variable example based upon data rejection. Phys. Rev. D **2**(8), 1418–1425 (1970)
43. Clauser, J.F., Horne, M.A.: Experimental consequences of objective local theories. Phys. Rev. D **10**(2), 526–535 (1974)
44. Brunner, N., Cavalcanti, D., Pironio, S., Scarani, V., Wehner, S.: Bell nonlocality. Rev. Mod. Phys. **86**, 419–478 (2014)
45. Giustina, M., et al.: Bell violation using entangled photons without the fair-sampling assumption. Nature **497**, 227–230 (2013)
46. Christensen, B.G., et al.: Detection-loophole-free test of quantum nonlocality, and applications. Phys. Rev. Lett. **111**, 130406 (2013)
47. Eberhard, P.H.: Background level and counter efficiencies required for a loophole-free Einstein–Podolsky–Rosen experiment. Phys. Rev. A **47**, 747–750 (1993)
48. Marcum, J.I.: Table of Q functions, Technical report, RAND Corporation, Santa Monica (1950) U.S. AirForce RAND Research Memorandum M-339
49. Abramowitz, M., Stegun, I.: Handbook of Mathematical Functions with Formulas, Graphs, and Mathematical Tables. Dover Publications, New York (1964)